# Optimizing Nurse Scheduling: A Supply Chain Approach for Healthcare Institutions

[1]**Jubin Thomas**

[1]Independent Researcher, Media, Pennsylvania, USA, Email: jubin.thomas@ieee.org

**Abstract**

When managing an organization, planners often encounter numerous challenging scenarios. In such instances, relying solely on intuition or managerial experience may not suffice, necessitating a quantitative approach. This demand is further accentuated in the era of big data, where the sheer scale and complexity of constraints pose significant challenges. Therefore, the aim of this study is to provide a foundational framework for addressing personnel scheduling, a critical issue in organizational management. Specifically, we focus on optimizing shift assignments for staff, a task fraught with complexities due to factors such as contractual obligations and mandated rest periods. Moreover, the current landscape is characterized by frequent employee shortages across various industries, with many organizations lacking efficient and dependable management tools to address them. Therefore, our attention is particularly drawn to the nurse rostering problem, a personnel scheduling challenge prevalent in healthcare settings. These issues are characterized by a multitude of variables, given that a single healthcare facility may employ hundreds of nurses, alongside stringent constraints such as the need for adequate staffing levels and rest period's—post-night shifts. Furthermore, the ongoing COVID-19 pandemic has exacerbated staffing challenges in healthcare institutions, underlining the importance of accurately assessing staffing needs and optimizing shift allocations for effective operation amidst crisis situations.

## 1. Introduction

In the realm of organization management, planners are frequently confronted with a myriad of complex and multifaceted challenges. These challenges span a wide spectrum, ranging from resource allocation and scheduling dilemmas to strategic decision-making processes. Traditionally, planners relied heavily on qualitative methods, drawing upon common sense and past management experiences to navigate these intricacies. However, with the advent of the digital age and the proliferation of big data—there has been a paradigm shift towards a more quantitative approach in addressing organizational issues [1]–[3]. The transition towards quantitative methodologies has become increasingly imperative in light of the magnitude and complexity of contemporary organizational problems. In today's fast-paced and data-driven environment, decision-makers are inundated with vast amounts of information, necessitating sophisticated analytical techniques to extract actionable insights [4]–[16]. Moreover, the constraints within which planners operate have become increasingly intricate and stringent—further underscoring the need for a systematic and rigorous approach to problem-solving. Therefore, at the heart of this paradigm shift lies the recognition that conventional methods alone are no longer sufficient to tackle the intricate challenges faced by organizations. Whether it is optimizing resource utilization, improving operational efficiency, or enhancing customer satisfaction—a quantitative approach offers a systematic framework for addressing these issues effectively. However, by leveraging advanced analytical tools and mathematical models, planners can gain deeper insights into complex organizational dynamics, enabling them to make more informed and data-driven decisions.

Against this backdrop, the objective of this study is to develop a fundamental framework for addressing a key issue within the realm of organization management—personnel scheduling. Personnel scheduling, particularly in large-scale organizations with diverse workforce requirements, presents a formidable challenge. The task of efficiently allocating shifts to staff members is fraught with complexities, ranging from managing the sheer number of employees to complying with contractual obligations and regulatory mandates. One specific subset of personnel scheduling concerns that this study aims to address is the nurse rostering problem, which is particularly prevalent in healthcare settings. Healthcare facilities, such as hospitals, operate in highly dynamic environments characterized by fluctuating patient volumes, diverse patient care needs, and stringent regulatory requirements. Consequently, scheduling nursing staff presents unique challenges due to the intricate interplay of numerous variables, including staffing levels, skill mix requirements, and regulatory constraints. Furthermore, healthcare facilities often grapple with staffing shortages, especially during crises such as the ongoing Covid-19 pandemic. In such critical situations, the need for efficient and effective personnel scheduling becomes even more pronounced. Schedulers must not only optimize shift allocations to meet patient care needs but also anticipate and mitigate staffing shortages by identifying the need for supplemental nursing staff and determining the required quantity.

The paper is structured into several sections, each contributing to a comprehensive understanding of the research topic. The background section provides a contextual overview of the study in Section 2, delving into the complexities of personnel scheduling in organizational management, with a specific focus on the nurse rostering problem in healthcare settings. In the Section 3, the paper reviews existing literature and research studies pertinent to personnel scheduling and optimization techniques. Section 4 outlines the approach taken to address the research objectives, including the use of quantitative methodologies and optimization techniques to develop a comprehensive framework for personnel scheduling. Section 5 presents the results of empirical studies conducted to validate the proposed framework. The result analysis section further examines the findings from the experimental studies, delving into specific metrics and performance indicators to evaluate the performance of the proposed framework in Section 6. Finally, the conclusion and future works section summarizes the key findings of the study and offers insights into potential avenues for future research in Section 7.

## 2. Background

In this section, we focus on mathematically representing personnel scheduling challenges, particularly the nurse rostering problem, within the framework of Integer Linear Programming (ILP). We emphasize the flexibility of this approach, acknowledging the need for model adaptations to suit different contexts. Before delving into nurse rostering problem-specific models, we establish a foundational understanding of essential components in personnel scheduling frameworks, setting the stage for discussions on decision variables and objective functions. Consequently, nuanced factors like overtime hours, the potential inclusion of night shifts, and variations in workers' qualifications are not factored into this rudimentary formulation. These aspects, though crucial, are intentionally omitted to maintain the broad applicability of the model. The time frame for this study is partitioned into distinct weeks, denoted as $T$, such that $T$ can be expressed as the union of subsets $T_1$ to $T_q$, where $q$ represents the total number of weeks. Additionally, it's essential to ensure that the time horizon is divisible into weeks due to the nature of defining working hours on a weekly basis. Within this context, the parameters $h_{min}$ and $h_{max}$ represent the minimum and maximum weekly working hours, respectively. For each shift $s$ belonging to the set $S$ and day $t$ within the time horizon $T$, the variable $a_{s,t}$ signifies the required number of workers during that particular shift. Moreover, the duration of each shift is represented by the parameter $d_s$. Furthermore, the variable $x_{i,s,t}$ is defined to indicate whether the $i^{th}$ worker opts to take the shifts $s$ on day $t$. Specifically, $x_{i,s,t}$ equals 1 if the $i^{th}$ worker chooses to undertake the shifts s on day $t$; otherwise, it is 0. Note that set $T$ is partitioned into groups, therefore $T = T_1 \cup \ldots \cup T_q \wedge T_h \cap T_k = \emptyset \; for \; h \neq k$, representing the weeks into which the time horizon is divided.

In considering optimal workplace dynamics, another aspect deserving attention is the equilibrium in working hours among employees. Analogous to the rationale presented earlier, an objective function can be devised with the aim of equalizing the distribution of working hours across the workforce. This approach seeks to mitigate discrepancies in workload allocation, fostering a more equitable environment within the organizational framework. In the context of work scheduling, various constraints govern the allocation of shifts to workers. One fundamental constraint, denoted by Equation (1), mandates that each worker is assigned to at most one shift per day. This constraint ensures that employees do not overlap their assignments, thereby maintaining a coherent schedule. The subsequent constraints, i.e. Equations (2) and (3), delineate the permissible bounds on weekly working hours for each worker. These bounds, specified as $h_{min}$ and $h_{max}$ respectively, play a crucial role in managing the workload distribution and ensuring compliance with labor regulations. Another critical aspect addressed by Equation (2.6) pertains to the coverage requirement for shifts. This constraint stipulates that there must be a minimum number of employees available to cover each shift adequately. By enforcing this requirement, the scheduling system ensures sufficient workforce availability to meet operational demands across different time periods. These constraints collectively form the backbone of an effective work scheduling system, balancing the needs of the organization with the well-being of its workforce.

$$\sum_{s \in S} x_{i,s,t} \leq 1; i \in I, t \in T \tag{1}$$

$$\sum_{t \in T_k} \sum_{s \in S} d_s x_{i,s,t} \geq h_{min}; i \in I, k = 1, \ldots q \tag{2}$$

$$\sum_{t \in T_k} \sum_{s \in S} d_s x_{i,s,t} \leq h_{max}; i \in I, k = 1, \ldots q \tag{3}$$

$$\sum_{i \in I} x_{i,s,t} \geq a_{s,t}; s \in S, t \in T \tag{4}$$

However, in personnel scheduling, achieving balanced work shifts is crucial. The goal is to evenly distribute shifts among workers while considering the challenge of assigning unequal shifts. Typically, this involves minimizing the workload of the most burdened worker, reflected in the objective function as minimizing the maximum workload. Balancing working hours is also essential, mirroring workload balance strategies. Constraints ensure fairness and efficiency, such as limiting workers to one shift per day and enforcing weekly working hour limits. Night shifts include mandatory rest periods, and weekly overtime is capped. This flexible formulation serves as the foundation for various scheduling models, enabling sophisticated solutions beyond human capacity. Extending the basic model involves considerations such as incorporating night shifts, handling overtime, and accommodating personalized preferences. Introducing worker satisfaction parameters allows for a nuanced optimization approach, wherein worker dissatisfaction is minimized. This addition augments the objective function, reflecting a comprehensive approach to workforce management. Additionally, the model can be adapted to address specific organizational needs, such as managing holiday periods, unplanned absences, or identifying staffing deficiencies. Applying these principles to nurse rostering problems, we define parameters tailored to the healthcare context. Parameters such as shift types, days, and nurse sets form the basis of the scheduling framework. The objective functions are customized to prioritize workload distribution and minimize overtime, aligning with healthcare operational goals. The formulation incorporates constraints specific to nursing practice, ensuring adequate staffing levels and compliance with regulatory requirements.

## 3. State-of-the-Art

The literature extensively covers the problem modeling and various methodologies concerning scheduling. Two prevalent scheduling types emerge—cyclical [17] and non-cyclical [18]. Cyclical scheduling entails the repetition of the same schedule until requirements alter. While straightforward to construct, these schedules might lack adaptability to changes such as [19], [20]. Conversely, non-cyclical scheduling involves generating a fresh schedule for each period, allowing greater flexibility to accommodate demand fluctuations such as [21], [22]. Optimization methods, such as linear, integer, or mixed integer programming, have been employed to address the nurse scheduling problem in various studies such as [23], [24]. Additionally, goal programming and constraint programming have been utilized in [25]. Recent studies have increasingly turned to metaheuristic techniques like genetic algorithms [26], tabu search [27], and simulated annealing [28]. We posit that solver-based resolution techniques hold promise for applicability in hospital services due to their transferability. Conversely, heuristics or metaheuristics may pose challenges in accessibility and time consumption such as [26]. Our contribution aims at enhancing linear programming solutions, optimizing demand coverage while minimizing salary costs, and maximizing nurse preferences and team balance.

## 4. Methodology

Despite the theoretical possibility for a single nurse to extend their work hours infinitely, as there are no inherent restrictions on overtime, practical considerations outlined in Equations (1) and (5) necessitate mandatory rest shifts. Consequently, in scenarios characterized by heightened patient influx, the requirements stipulated by these constraints may clash with Equation (6), rendering the problem formulation untenable. This situation manifests as understaffing within the healthcare department, prompting the need for an assessment regarding the necessity of recruiting additional personnel and determining the optimal number of new hires. Therefore, in our investigation, the primary inquiry revolves around the feasibility of resolving the given issue. To commence this examination, we introduce a novel variable denoted as $j_{s,t}$ from the set of natural numbers, denoted by $N$. This variable will be integrated into both the objective function and in Equation (6) for comprehensive analysis.

$$\sum_{s \in S} x_{i,s,t+1} \leq 1 - x_{i,night,t}; i \in I, t \in T \quad (5)$$

$$a_s \sum_{i \in I} x_{i,s,t} \geq d_t; s \in S, t \in T \quad (6)$$

However, Equation (7) reveals that it does not render the problem inadmissible. If there is a shortage of nurses to attend to all patients, a positive value would be assigned to the variable $j_{s,t}$, representing the hypothetical number of auxiliary nurses assigned to shift $s$ of day $t$. While these "dummy" nurses are not real entities and are not featured in any other constraints, their inclusion ensures the problem's admissibility from the outset. To prevent the solver from opting for an allocation strategy where all shifts are covered solely by fictitious nurses, the variable $j_{s,t}$ is incorporated into the Equation (8) with a significantly high weight, typically denoted by the parameter $M$ and commonly set at $10^6$.

$$a_s(\sum_{i \in 1} x_{i,s,t} + j_{s,t}) \geq d_t; s \in S, t \in T \quad (7)$$

$$\min z + p_1 \sum_{i \in I} \sum_{k=1}^{q} \beta_{i,k} + M \sum_{s \in S} \sum_{t \in T} j_{s,t} \quad (8)$$

Therefore, in our study, the solver is tasked with determining whether to allocate a non-zero value to a variable based on the solvability of the basic problem. This approach serves the purpose of identifying instances of understaffing and conveying this information to the decision maker. This means that the solver will decide to assign a non-zero value to the variable only if the basic problem does not admit a solution. We therefore use the values assumed by the variable only to identify situations of understaffing and to communicate them to the decision maker. In situations where the current staffing is inadequate, necessitating the recruitment of additional personnel, it becomes imperative to resolve the problem afresh. Consequently, determining the ideal number of nurses to employ is crucial. This decision must consider not only the initial hiring costs but also the subsequent impact on workload distribution and reduction in overtime expenses. However, the integration of new nurses into the existing model poses a challenge as it necessitates a re-evaluation and recalibration of the entire system. For each additional nurse, the model undergoes modification, demanding a comprehensive reiteration of the problem-solving process. To address this challenge, we propose the pseudo code as shown in Fig. 1. Based on the comparison, the algorithm determines whether hiring additional nurses is the optimal solution, considering both the immediate recruitment costs and the long-term benefits of workload redistribution and reduced overtime expenses The influence of the parameter '$c$' on the algorithm's progression is evident. A very low or zero value of '$c$' poses the risk of updating the set '$I$' an infinite number of times. This scenario arises because the addition of a new nurse invariably leads to a more balanced workload distribution, resulting in reduced overtime hours and, consequently, a decrease in the objective function's value. However, this potential issue is mitigated by the presence of Equation (5), which mandates a minimum number of hours for each nurse's workload. This ensures that the algorithm, as described earlier, converges to a solution within a finite number of steps.

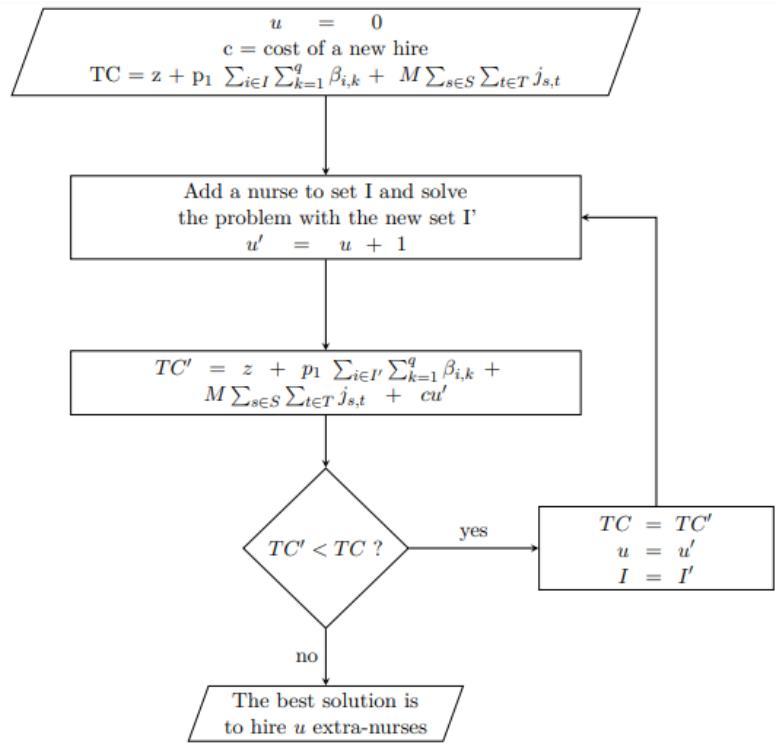

Fig. 1. Flowchart of the pseudo code

## 5. Experimental Analysis

In this section, we delve into the practical application of the theoretical concepts—specifically, we focus on the Python code, executed within Jupyter notebooks, designed to tackle the nurse rostering problem. Our objective is to develop an efficient solution methodology by leveraging essential input data, including employee details, reference time frame, and daily patient volume. The Python code serves as a computational framework aimed at optimizing the nurse rostering problem, aiming to streamline nurse scheduling processes within healthcare facilities. Through meticulous implementation, we aim to address the intricate scheduling challenges prevalent in healthcare settings, ultimately enhancing operational efficiency and patient care quality. In establishing our model, we employed the Pyomo[1] modeling language, which is an open-source software package built on Python. Pyomo facilitated the formulation of our optimization problem, allowing us to define variables, an objective function, and constraints effectively. Subsequently, we utilized IBM ILOG CPLEX V12.9.0[2] as the solver, known for its efficiency in handling complex optimization tasks, to determine the optimal solution within the specified constraints. In addition to Pyomo and CPLEX, we leveraged other open-source libraries such as Pandas and NumPy. These libraries played a crucial role in data manipulation and numerical computing, augmenting our modeling efforts and enhancing the overall efficiency of our solution approach. By combining these tools, we were able to develop a comprehensive and effective solution methodology for the personnel scheduling problem. The initial dataset, crucial for configuring the model, has been imported utilizing the pandas library from dedicated Excel spreadsheets. For instance, a typical corporate database may resemble the structure utilized in this study, focusing primarily on row designations rather than additional attributes, which were included solely for illustrative purposes. Additionally, another pertinent dataset pertains to the daily patient count within the ward. While contemplating potential enhancements to the model, one might consider augmenting this dataset with supplementary attributes, such as the minimum nursing staff required per day. It's notable that instead of representing the days as numerical sequences, which would simplify constraint definitions, we chose to depict them as strings. Consequently, this necessitates the establishment of the 'day_after' dictionary, as delineated in the provided code snippet. Certain parameters, such as shift durations, weekly working hour limits, have been determined arbitrarily, with an aim to maintain a semblance of realism. We introduced the $w_{s,t}$ parameter, which plays a crucial role in our model. Defining scientific criteria for determining the "heaviness" of each shift poses a challenge. Therefore, the weights assigned to the combinations of shifts ($s$, $t$) are arbitrary. For instance, we considered night shifts to be heavier than day shifts and weekend shifts to be heavier than midweek shifts. Consequently, the shifts with the highest weight are the night shifts during the weekends. However, it is plausible to envision a customization of these weights without compromising the model's functionality. For customization, nurses themselves could provide weights, perhaps on a scale from 1 to 100, for each shift based on their experiences and preferences. Alternatively, nurses could express their approval ratings for different shifts. Another approach involves tailoring the workload according to the age of the nurses. Moreover, this observation underscores the intricate nature of workload assessment and its profound impact on practical decision-making processes. The complexity involved in determining these workload parameters highlights the importance of thorough analysis and consideration in real-world applications. By understanding and leveraging such insights, decision-makers can devise more tailored and effective strategies to optimize personnel scheduling and enhance overall operational efficiency.

---

[1] http://www.pyomo.org/
[2] https://www.ibm.com/support/pages/downloading-ibm-ilog-cplex-enterprise-server-v1290

## 6. Result Analysis

In this section, our focus lies on assessing the effectiveness of the model concerning its ability to efficiently identify optimal solutions and the quality of those solutions. Our methodology involves testing the model across various datasets, each differing in the number of variables, thereby impacting the overall problem size. The Table 1 illustrates the time required by the solver, employing the branch and cut algorithm, to discover a solution corresponding to different problem sizes, characterized by the number of nurses ($n$) and the number of weeks ($q$). As the problem size increases, reflected in larger values of $n$ and $q$, there is a corresponding slight increase in the time required to reach a solution. Specifically, for smaller problem instances with 10 nurses and 4 weeks, the solver completes its task in 0.32 seconds. With an increase in the number of nurses to 40 while keeping the number of weeks at 4, the time taken rises marginally to 0.34 seconds. Similarly, for larger problem sizes with 40 nurses and 8 weeks, the solver requires 0.49 seconds to find a solution. These findings provide insights into the computational efficiency of the model across different problem configurations.

Table 1. Relationship between the problem's dimension and its resolution time

| $n$ | $q$ | time ($s$) |
| --- | --- | --- |
| 10 | 4 | 0.32 |
| 20 | 4 | 0.34 |
| 20 | 6 | 0.42 |
| 40 | 6 | 0.47 |
| 40 | 8 | 0.49 |

However, in assessing a feasible solution obtained through optimization, it is imperative to gauge its effectiveness through specific Key Performance Indicators (KPIs). These KPIs offer insights into various aspects of the solution's performance. Firstly, the minimum and maximum workload across all tasks and shifts are evaluated. This involves summing the workload for each task $t$ under shift $s$ and department $i$, and identifying the minimum and maximum values across all departments. Additionally, the minimum and maximum weekly hours for each department $i$ over all weeks $k$ are calculated. This computation considers the labor standards and the assigned workload for each task within the specified weeks. Moreover, the total overtime hours incurred by each department are determined by summing the overtime coefficients across all weeks $k$ for each department $i$. It is essential to note that these performance indicators serve not to provide absolute judgments but rather to facilitate comparative analyses among different solutions. By scrutinizing these KPIs, decision-makers can effectively evaluate and compare the efficacy of various solutions in addressing workforce scheduling challenges. It determines the optimal solution, suggesting the hiring of only one additional nurse as shown in Fig. 2. Subsequently, an analysis of KPIs reveals significant improvements following this decision. Notably, the reduction in overtime expenditure surpasses the incurred cost of hiring a new nurse. Consequently, the objective function's value decreases, indicating enhanced efficiency. This instance underscores the pivotal role of monitoring and assessing performance indicators in understanding model evolution and aiding decision-making processes.

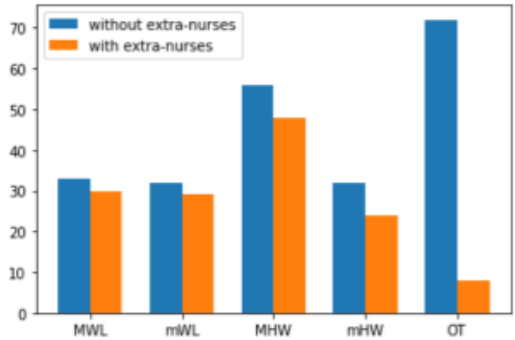

Fig. 2. KPI variations between the two ideal solutions

Once the staff scheduling has been completed, it's vital to share the outcomes with both management and personnel. We've discussed how overall performance indicators can inform management decisions and notify the management about staffing needs. However, it's equally important to communicate the results in a non-technical manner. One approach is to generate an Excel workbook as part of the code output. Each nurse has a dedicated sheet, automatically populated with their shifts using Python's Openpyxl[3] library. This format allows for quick visual assessment of scheduling accuracy, ensuring each nurse has a maximum of one shift per day and adequate rest after night shifts. However, this method may lead to longer code execution times, especially for larger scheduling tasks involving many nurses over extended periods. As the Excel document grows in size, navigating through it could become challenging. An alternative to streamline code execution and simplify result interpretation is to utilize the drop-down tool provided by the Ipywidgets[4] library. Users can select specific nurses or view daily duty assignments with ease. Following the optimization algorithm's execution for determining new hiring needs, the drop-down menu updates to reflect any additional staff. However, we observed the significance of determining the optimal number of new nurses to employ, which involves striking a balance between the cost implications of new hires and those associated with overtime hours, alongside the subsequent impact on the objective function due to workload redistribution. It's evident that this equilibrium is heavily influenced by the values assigned to parameters $p_1$ and $c$. For the planner, understanding how the solution fluctuates with varying parameter values is of considerable interest. To facilitate this, we have developed an additional interactive tool as shown in Fig. 3. Through this tool, users can dynamically adjust the values attributed to parameters $p_1$ and $c$, enabling them to visualize corresponding alterations in the solution as well as key KPIs.

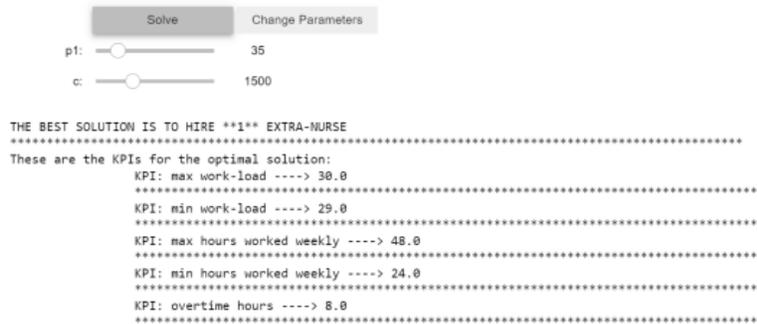

Fig. 3. Interactive changes to the $p_1$ and $c$ parameters

## 7. Conclusion and Future Works

In this study, we delve into the mathematical modeling of intricate scenarios, particularly those associated with nurse rostering problem, aiming for prompt insights aiding managerial decisions grounded in empirical evidence. The developed formulation, tailored to the healthcare setting, exhibits versatility across diverse contexts, with potentially fewer constraints than our specific case. For instance, scenarios devoid of nocturnal operations can also benefit from efficient staff scheduling practices, echoing the imperative for adaptable solutions while upholding precision and thoroughness. Our model embodies a paramount attribute— flexibility. This pivotal characteristic enables its applicability across varied contexts without compromising on the meticulousness demanded by complex decision-making processes. While acknowledging the utility of existing tools, especially in specialized domains, the necessity for advanced methodologies becomes evident when confronting substantial problem scales and rigid constraints. Such demand is particularly pronounced in large-scale enterprises, where effective workforce management stands as a pivotal determinant of operational success. The adverse effects of demanding shifts on employee motivation and performance underscore the significance of optimal scheduling practices, resonating strongly within competitive job markets. Therefore, the extensibility of our model is notable, offering avenues for augmentation across multiple dimensions. Expansion possibilities encompass broader temporal considerations, potentially spanning six-month or annual horizons, integrating constraints associated with employee leave periods. Moreover, augmenting shifts to include supervisory roles or specialized skill requirements, alongside accommodating trainees with tailored responsibilities, presents further avenues for refinement. However, our investigative focus remains directed towards critical aspects—overtime management, addressing staffing shortages through recruitment strategies, and the equitable distribution of workloads.

---

[3] https://pypi.org/project/openpyxl/
[4] https://pypi.org/project/ipywidgets/